\documentclass[11pt]{article}
\usepackage{geometry}                % See geometry.pdf to learn the layout options. There are lots.
\usepackage{amsmath}
\usepackage{amscd}
\usepackage{times}
\usepackage{epsfig}
\usepackage{amsfonts,amsthm,eucal,amssymb,amscd}
\usepackage{color}
\usepackage{graphicx}
\usepackage{epstopdf}
\usepackage{authblk}
\title{Superintegrable Lissajous systems on the sphere}

\author[1]{J.A. Calzada}
\author[2]{\c{S}. Kuru}
\author[3]{J. Negro}
\affil[1]{Departamento Matem\'atica Aplicada, Universidad de
Valladolid, 47011 Valladolid, Spain} \affil[2]{Department of
Physics, Faculty of Sciences, Ankara University, 06100 Ankara,
Turkey} \affil[3]{Departamento de F\'{\i}sica Te\'orica, At\'omica y
\'Optica, Universidad de Valladolid\\47071 Valladolid, Spain}
\begin{document}
\maketitle

\begin{abstract}
A kind of systems on the sphere, whose trajectories are similar to the Lissajous
curves, are studied by means of one example.
The symmetries are constructed following a unified and
straightforward procedure for both  the quantum and the classical versions of the model.
In the quantum case it is stressed how the symmetries give the degeneracy of
each energy level. In the classical case it is shown how the constants of
motion supply the orbits, the motion and the frequencies in a natural way.
\end{abstract}

%\begin{keyword}

%Superintegrable system,  Lissajous curve,
%ladder operator, shift operator, trajectory, constant of motion.

%\PACS 03.65.Fd,02.20.Sv, 02.30.Ik
%\end{keyword}
%\end{frontmatter}

%%%%%%%%%%%%%%%%%%%%%%%%%
\section{Introduction}

The well known Lissajous curves in two dimensions (2D) correspond to the
motion of a system that can be described by two cartesian coordinates,
where each coordinate is harmonic in time and the ratio of both
frequencies is a rational number. Such curves are closed in a rectangle
and depend on the phase difference at the initial time.
In a similar way, we want to extend this
point of view to systems on the sphere where the motion is obtained
in terms of the  spherical
coordinates $\varphi(t)$ and $\theta(t)$. In this case, the motion in each coordinate
will be periodic but not harmonic, and their trajectories closed curves inscribed in
a `spherical rectangle', $\varphi_1\leq\varphi\leq\varphi_2$, $\theta_1\leq\theta\leq\theta_2$,
when the proportion of the two periods is a rational number.

The type of systems leading to Lissajous type trajectories include the TTW \cite{ttw09,ttw10}
and other similar systems \cite{kalnins10,kalninskress10,post10,kalnins12,ranada12,ranada13,post}.
They are superintegrable  and the search of their symmetries has been the subject of a
considerable number of recent contributions.
In this work, by means of one example, we want to stress the features of such systems
defined on the sphere leading to Lissajous like curves. In particular
it will be shown how the trajectories are determined and how the period of the
motion in each coordinate can be computed by means of pure algebraic methods.
We will adopt a simple procedure presented in \cite{celeghini13} in order to deal with
the symmetries and constants of motion.

Let us emphasize the main novelties of our approach:
(i) The method to find the symmetries for both the quantum and the classical versions of these
systems follow the same pattern. In previous references, quite different procedures
were applied in order to obtain classical or quantum symmetries, as a consequence,
the origin of the close relationship of  classical and quantum algebraic relations was hidden from the very beginning.
 (ii) Our way relies on simple arguments based
on the factorization properties of one--dimensional systems
\cite{schrodinger40,infeldhull51,sengul08}. This method, well known
in quantum mechanics, is extended to classical systems. For
instance, we can define in a straightforward way `ladder functions'
and `shift functions' whose counterpart operators are familiar in
quantum mechanics. (iii) As we will see later, the classical version
of our 2D system is maximally superintegrable, so that we can find
three independent constants of motion (with two of them in
involution). These constants will characterize the orbits. However,
one additional constant of motion, depending explicitly on time, is
obtained from some `ladder functions'. This additional constant of
motion will determine the motion of the system along each orbit.

In summary, we have a kind of `complete superintegrability'
where the number of independent integrals of motion is `$(2n-1) + 1=2n$'
characterizing completely the motion (not only the trajectory). The symmetries
or constants of motion in this paper include square roots that, in the case
of operators, can only be defined when they act on eigenfunctions. Hence,
the symmetries are not polynomial in the momentum operators. However they can
be translated into polynomial symmetries in a simple way, as it will be shown
in a forthcoming work \cite{polynomial14}.

The paper is organized as follows. Section 2 is devoted to the
quantum version of our example. This system can be considered as a
composition of two one--dimensional trigonometric P\"oschl--Teller
(PT) potentials. Therefore, the well known factorization properties
of the PT component systems can be straightforwardly applied in
order to get the symmetries of the composed system. It will be shown
how these symmetries explain the degeneracy of each
energy level. Section 3 will be dedicated to the classical system
whose constants of motion are found by implementing the methods used
previously in the quantum case. These symmetries will determine the
trajectories and their properties as Lissajous curves. Section 4
will be concerned with the motion of the system. In this case, the
motion can also be  obtained in an algebraic way by means of the
aforementioned additional constant of motion depending explicitly on
time. In particular, the frequency takes part of the algebraic
properties through Poisson brackets.  Some comments and remarks in
Section 5 on the special character of Lissajous systems will end the
paper.

%%%%%%%%%%%%%%%%%%%%%%
\section{The quantum  system}

The Hamiltonian operator that we will consider in this paper belongs
to a type of Smorodinsky--Winternitz systems on the sphere
\cite{ballesteros,ballesteros13}. It depends on a real coupling
parameter, $k$, and has the following form (the units $2m=1$ and
$\hbar=1$ have been chosen to simplify the formulas):
\begin{equation}\label{qh1}
\hat{H} = - \partial^2_{\theta} - \cot{\theta}\partial_{\theta}
-\frac{1}{\sin^2 {\theta}}
\partial^2_{\varphi}-\frac{k^2}{\sin^2
{\theta}}\left(\frac{1/4-\alpha^2}{\cos^2{k\,\varphi}}
+\frac{1/4-\beta^2}{\sin^2{k\,\varphi}}\right)
\end{equation}
where $0<\theta<\pi$ and  $0<\varphi<\pi/2k$. The first three terms correspond
to the Laplacian in spherical coordinates, the last term is for the potential.
It should be remarked that $k$ must satisfy $k\geq1/4$, so that the system be well
defined in a region of the sphere.

First of all, the change of variable ${k\,\varphi}=\phi$ will be performed, so that
henceforth we will work with the following Hamiltonian \cite{celeghini13}
\begin{equation}\label{qh2}
\hat{H} = - \partial^2_{\theta} - \cot{\theta}\partial_{\theta}
+\frac{k^2}{\sin^2 {\theta}}\left(
-\partial^2_{\phi}+\frac{\alpha^2-1/4}{\cos^2{\phi}}+\frac{\beta^2-1/4}{\sin^2{\phi}}\right)\,,
\end{equation}
where now, the angle $\phi$ will range in the interval $(0,\pi/2)$.
%All the properties such as superintegrability, constants of motion, eigenfunctions,
%etc. of this Hamiltonian are inhered by the initial one. The change of the azimuthal
%variable consists in a dilation.
The corresponding eigenvalue equation is
\begin{equation}\label{eve1}
\hat{H}\,\Psi(\theta,\phi)=E\,\Psi(\theta,\phi)\,.
\end{equation}
It is clear that the Hamiltonian (\ref{qh2}) is separated in
the spherical coordinates $(\theta,\phi)$, so we will look for separable solutions,
$\Psi(\theta,\phi)=\Theta(\theta)\,\Phi(\phi)$. This type of solutions are characterized by
\begin{equation}\label{eve2}
\hat{H}^M_{\theta}\,\Theta(\theta)=E_{\theta}\,\Theta(\theta),
\quad\hat{H}_{\phi}\,\Phi(\phi)=E_{\phi}\,\Phi(\phi),
%\quad E_\theta:= E, E_{\phi}:= \epsilon^2\,
\end{equation}
where the one--dimensional Hamiltonians are
\begin{equation}\label{qht1}
\hat{H}^M_{\theta} =
%- \partial^2_{\theta} -
%\cot{\theta}\partial_{\theta} +\frac{k^2\, E_{\phi}}{\sin^2 {\theta}}\,
%\equiv
- \partial^2_{\theta} -
\cot{\theta}\partial_{\theta} +\frac{M^2}{\sin^2
{\theta}}\,,
%\qquad M^2 := k^2 E_\phi = k^2\epsilon^2
\end{equation}
which has a form equivalent to a one--parameter trigonometric PT Hamiltonian, and
\begin{equation}\label{qhp2}
\hat{H}_{\phi} = - \partial^2_{\phi} +\frac{\alpha^2-1/4}{\cos^2
{\phi}}+\frac{\beta^2-1/4}{\sin^2{\phi}}
\end{equation}
which is a two--parameter trigonometric PT potential.
The following notation has been introduced
\begin{equation}\label{not}
E_\theta:= E\,,\qquad  E_{\phi}:= \epsilon^2\, ,\qquad
M^2 := k^2 E_\phi = k^2\epsilon^2 \,.
\end{equation}
The case $\beta=\pm 1/2$ (or $\alpha=\pm 1/2$) in (\ref{qhp2}) corresponds
to the one--parameter trigonometric PT potential (in this case,
$\phi\in (-\pi/2,\pi/2)$ and $k\geq 1/2$).
%It will be useful to express the  eigenvalue $E_{\phi}$ in terms of $\epsilon$,
%\begin{equation}\label{energia}
%E_{\phi}=\epsilon^2
%\end{equation}
%so that $\hat{H}_{\theta}$ acting on the eigenfunctions (\ref{eve2}) can be rewritten as
%\begin{equation}\label{qht2}
%\hat{H}_{\theta} = - \partial^2_{\theta} -
%\cot{\theta}\partial_{\theta} +\frac{(k\, \epsilon)^2}{\sin^2 {\theta}}
%\end{equation}
%and the eigenvalues of $\hat H$ for these eigenfunctions correspond to $E= E_\theta$.

In order to get the symmetries of $\hat H$, we will deal separately
with each of these two one--dimensional problems by means of
factorizations \cite{infeldhull51}. We will find   ladder (lowering
and raising) operators for $\hat{H}_{\phi}$ and   shift operators
for $\hat{H}_{\theta}$. As we will see later on, the ladder
operators will act on the coefficient $k\epsilon$ of the Hamiltonian
(\ref{qht1}) for the two--parameter PT potential in the form (for
the one--parameter PT potential the action is slightly different)
\begin{equation}\label{actions2}
k\epsilon\to k(\epsilon \pm 2)
\end{equation}
while in the case of the shift operators their action is
\begin{equation}\label{actions2}
 k \epsilon \to k \epsilon \pm 1\,.
\end{equation}
%these two kind of operators will have the following action
%on the coefficient $k\epsilon$ of the Hamiltonian (\ref{qht1}), respectively:
%\begin{equation}\label{actions2}
%k\epsilon\to k(\epsilon \pm 2),\qquad k \epsilon \to k \epsilon \pm
%1\,.
%\end{equation}
%In the case where instead of (\ref{qhp2}) we have a one--parameter PT Hamiltonian,
%\begin{equation}\label{qhp1}
%\hat{H}_{\phi} = - \partial^2_{\phi} +\frac{\alpha^2-1/4}%{\cos^2
%{\phi}}
%\end{equation}
%the ladder operators have a slightly different
%effect than in the previous two--parameter case:
%\begin{equation}\label{actions1}
%k\epsilon\to k(\epsilon\pm 1)
%\end{equation}
Then, the symmetries of the
Hamiltonian (\ref{qh2}) are found by combining these two actions
corresponding to ladder and shift operators
in order to keep invariant the value of $k\epsilon$. The details will be given
in Section 2.4.
%%%%%%%%%%%%%%%%%%%%%%%%%%%%%%%%%%%%

\subsection{Ladder operators of the two--parameter P\"oschl--Teller Hamiltonian}
 The lowering and raising operators for the two--parameter PT Hamiltonian
$H_\phi$ (\ref{qhp2}) take the form \cite{infeldhull51,calzada12}
\begin{equation}
\begin{array}{l}
\hat{B}_{\epsilon}^+ =(\epsilon+1) \sin 2\phi\, \partial_\phi +
\epsilon(\epsilon+1) \cos 2\phi- \alpha^2 +  \beta^2\,,
\\[2.ex]
%+\epsilon
% (k-1)-\sqrt{4 \beta^2+k^2-1}+k(k+\sqrt{4 \beta^2+k^2-1}-1)
%\\[2.ex]
\hat{B}_{\epsilon}^-= -(\epsilon-1) \sin 2\phi \,\partial_\phi
 + \epsilon(\epsilon-1) \cos 2\phi- \alpha^2 +\beta^2\,.
%\\[2.ex]
%-\epsilon
% (k-1)-\sqrt{4 \beta^2+k^2-1}+k(k+\sqrt{4 \beta^2+k^2-1}-1)
\end{array}
\end{equation}
These ladder operators sometimes are called pure--ladder in order to
stress that they change only the energy of the system.  The action
on the eigenfunctions $\Phi_\epsilon$ is as follows
\begin{equation}\label{xipm}
%\begin{array}{l}
\hat{B}^+ \Phi_\epsilon:=\hat{B}_\epsilon^+ \Phi_\epsilon \propto
\Phi_{\epsilon+2}, \qquad \hat{B}^- \Phi_{\epsilon+2}
:=\hat{B}_\epsilon^- \Phi_{\epsilon+2} \propto \Phi_{\epsilon}\,
%\end{array}
\end{equation}
where, according to the notation (\ref{not}), formally the action of $\sqrt{\hat{H}_\phi}$ on
$\Phi_\epsilon(\phi)$ is
\begin{equation}\label{xipm2}
\sqrt{\hat{H}_\phi}\, \Phi_\epsilon(\phi)= \epsilon \, \Phi_\epsilon(\phi)\, .
\end{equation}
By using (\ref{xipm}) and (\ref{xipm2}) it is shown that the
free--index ladder operators $\hat{B}^\pm $ satisfy the commutation
relation
\begin{equation}\label{cxi}
[\sqrt{\hat{H}_\phi},\hat{B}^\pm ]=\pm 2 \,\hat{B}^\pm \, .
\end{equation}
The consecutive action of $\hat B^\pm$ can be expressed in the following
form that will be useful later
\begin{equation}\label{BBB}
\begin{array}{c}
(\hat{B}^-_{\epsilon-2n} \dots
\hat{B}^-_{\epsilon-4}\hat{B}^-_{\epsilon-2}) \sqrt{\hat{H}_\phi} =
(\sqrt{\hat{H}_\phi} + 2n)\,(\hat{B}^-_{\epsilon-2n}\dots \hat{B}^-_{\epsilon-4}\hat{B}^-_{\epsilon-2})\\[2.ex]
(\hat{B}^+_{\epsilon+2(n-1)} \dots
\hat{B}^+_{\epsilon+2}\hat{B}^+_{\epsilon}) \sqrt{\hat{H}_\theta} =
(\sqrt{\hat{H}_\phi} - 2n)\,(\hat{B}^+_{\epsilon+2(n-1)}\dots
\hat{B}^+_{\epsilon+2}\hat{B}^+_{\epsilon})
\end{array}\nonumber
\end{equation}
or, in a shorter notation
\begin{equation}\label{bh}
(\hat{B}^\pm)^n \sqrt{\hat{H}_\phi} = (\sqrt{\hat{H}_\phi} \mp
2n)\,(\hat{B}^\pm)^n,\qquad \forall n\in \mathbb N\,.
\end{equation}
The above relations (\ref{cxi}) and (\ref{bh}) are assumed to be
satisfied when we act on the eigenfunctions $\Phi_\epsilon$
of the Hamiltonian operator,
according to  (\ref{xipm}) and (\ref{xipm2}).

%%%%%%%%%%%%%%%%%%%%%%%%%%%%%%%%%%%%%%%%%%%%%%%%%%%%%%%%%%%%%%%%
\subsection{Ladder operators  for the one--parameter P\"oschl--Teller Hamiltonian}
The one--parameter PT Hamiltonian ($H_\phi$ with $\beta=1/2$) is
very well known, its  ladder operators are \cite{sengul09}
\begin{equation}\label{bpm}
\hat{B}_{\epsilon}^+=-\cos{\phi}\,\partial_{\phi}+\epsilon\,\sin{\phi},
\qquad\hat{B}_{\epsilon}^-=\cos{\phi}\,\partial_{\phi}+(\epsilon+1)\,\sin{\phi}
\end{equation}
and their action on %the normalizable
eigenfuncions is
\begin{equation}\label{bpmpsi}
\hat{B}^+\,\Phi_{\epsilon}:=\hat{B}_{\epsilon}^+\,\Phi_{\epsilon}\propto\Phi_{{\epsilon}+1},
\qquad\hat{B}^-\,\Phi_{\epsilon+1}:=\hat{B}_{\epsilon}^-\,\Phi_{\epsilon+1}\propto\Phi_{\epsilon}\,.
\end{equation}
They satisfy the following commutation relations
\begin{equation}\label{com1}
[\sqrt{\hat{H}_{\phi}},\hat{B}^{\pm}]=\pm\hat{B}^{\pm} \, ,
\end{equation}
or in other words,
\begin{equation}\label{com2}
\hat{B}^{\pm}\, \sqrt{\hat{H}_{\phi}}=(
\sqrt{\hat{H}_{\phi}}\mp1)\hat{B}^{\pm} \,,
\end{equation}
so, a similar relation to (\ref{bh}) will hold for the consecutive action
of $\hat{B}^{\pm}$.

%%%%%%%%%%%%%%%%%%%%%%%%%%%%%%%%%%%%%%%%%%%%%%%%%%%%%%%%%5
\subsection{Shift operators  for $\hat{H}^M_{\theta}$}
The one--dimensional Hamiltonian $\hat{H}_{\theta} ^M$ given in
(\ref{qht1}) is factorized in the following way \cite{sengul09}
\begin{equation}\label{facqht2}
\hat{H}_{\theta} ^M=\hat{A}^+_M\hat{A}^-_M+\lambda_M=\hat{A}^-_{M+1}\hat{A}^+_{M+1}+\lambda_{M+1},
\qquad\lambda_M=M(M-1)
\end{equation}
where
\begin{equation}\label{apm}
\hat{A}^+_M=-\partial_{\theta}+(M-1)\cot{\theta},
\qquad\hat{A}^-_M=\partial_{\theta}+M \cot{\theta}\,.
\end{equation}
The hierarchy of Hamiltonians (\ref{facqht2}) satisfy the following commutation rules
\begin{equation}\label{aint}
\hat{A}^-_M\hat{H}_{\theta} ^M=\hat{H}_{\theta} ^{M-1}\hat{A}^-_M,
\qquad\hat{A}^+_M\hat{H}_{\theta} ^{M-1}=\hat{H}_{\theta} ^M\hat{A}^+_M \, .
\end{equation}
We can write  these rules, in a shorter notation, by eliminating the $\hat{A}^\pm$
subindex, in the form
\begin{equation}\label{actionShift}
\hat{A}^-\hat{H}_{\theta} ^M=\hat{H}_{\theta} ^{M-1}\hat{A}^-,\qquad
\hat{A}^+\hat{H}_{\theta} ^{M-1}=\hat{H}_{\theta} ^M\hat{A}^+\, .
\end{equation}
The action on the eigenfunctions $\hat{H}_{\theta}^M \Theta^{M}_{E}= E\,\Theta^{M}_{E}$
of the Hamiltonian (\ref{qht1}) is
\begin{equation}\label{bpmpsi2}
\hat{A}^-\,\Theta^{M}_{E}:=\hat{A}^-_M\,\Theta^{M}_{E}\propto\Theta^{M-1}_{E},
\qquad\hat{A}^+\,\Theta^{M-1}_{E}:=\hat{A}^+_M\,\Theta^{M-1}_{E}\propto\Theta^{M}_{E}
\, .
\end{equation}
Therefore, this kind of operators keep the energy $E$, but change the parameter $M$,
this is the reason why they are called pure shift operators.
These operators satisfy the following commutation relations
\begin{equation}\label{coma1}
[\hat{H}_{\theta} ^M,\hat{A}_M^-]=
\frac{2M-1}{\sin^2{\theta}}\,\hat{A}_M^-,\qquad
[\hat{H}_{\theta} ^M,\hat{A}_M^+]=-\hat{A}_M^+
\,\frac{2M-1}{\sin^2{\theta}}\,.
\end{equation}

%%%%%%%%%%%%%%%%%%%%%%%%%%%%%%%%%%%%%%%%%%%%%%%%%%%%%%%%%55
\subsection{Symmetries of the Hamiltonian $\hat{H}$}
Now, we can construct the symmetry operators  $\hat{X}$ such that
\begin{equation}\label{symm}
[\hat{H},\hat{X}]=0\,.
\end{equation}
Combining the commutations (\ref{bh}) and (\ref{actionShift}), the
symmetry operators (for the two--parameter PT potential case) can be
constructed in the following way. Let us take
\begin{equation}\label{symmpm2}
\hat{X}^{\pm}=(\hat{A}^{\pm})^{2m}\,(\hat{B}^{\pm})^n
\end{equation}
where $m$ and $n$ are (positive) integer numbers.
Let us write the Hamiltonian (\ref{qh2}) in the following formal way
\begin{equation}\label{qht4}
\hat{H} \equiv \hat{H} \left[k\, \sqrt{\hat{H}_{\phi}}\right]= - \partial^2_{\theta} -
\cot{\theta}\partial_{\theta}
+\frac1 {\sin^2 {\theta}}\left(k\, \sqrt{\hat{H}_{\phi}}\right)^2
\end{equation}
in order to make explicit the dependence of $\hat{H}$ on the one--dimensional
Hamiltonian $\hat{H}_{\phi}$ (\ref{qhp2}).
Now, we can address the commutation of $\hat X^\pm$ and $\hat H$.
First, we have
\begin{equation}
(\hat{A}^{\pm})^{2m}\,(\hat{B}^{\pm})^n
\hat H\left[k\, \sqrt{\hat{H}_{\phi}}\right]=
(\hat{A}^{\pm})^{2m}\,\hat H\left[k( \sqrt{\hat{H}_{\phi}}\mp 2n) \right]
(\hat{B}^{\pm})^n
\end{equation}
where (\ref{bh}) has been applied.
%and replaced $\sqrt{\hat{H}_{\phi}}$ by its action
%$\epsilon$ on any eigenfunction $\Phi_\epsilon$.
Next, by means of (\ref{actionShift}),
\begin{equation}
(\hat{A}^{\pm})^{2m}\,\hat H\left[k( \sqrt{\hat{H}_{\phi}}\mp 2n) \right]
(\hat{B}^{\pm})^n=
\hat H\left[k( \sqrt{\hat{H}_{\phi}}\mp 2n) \pm 2m\right](\hat{A}^{\pm})^{2m}\,(\hat{B}^{\pm})^n
\,.
\end{equation}
%%
%Then, having in mind that $M= k\epsilon$, write the following product,
%\begin{equation}
%(\hat{A}^{\pm})^{2m}\,(\hat{B}^{\pm})^n \hat H_\theta^M=
%(\hat{A}^{\pm})^{2m}\,\hat H_\theta^{M'}(\hat{B}^{\pm})^n,\qquad
%M' = k(\epsilon \mp 2n)
%\end{equation}
%where we have applied (\ref{bh}) and replaced $\sqrt{\hat{H}_{\phi}}$ by its action
%$\epsilon$ on any eigenfunction $\Phi_\epsilon$.
%Next, by means of (\ref{actionShift}),
%\begin{equation}
%(\hat{A}^{\pm})^{2m}\,\hat H_\theta^{M'}(\hat{B}^{\pm})^n=
%\hat H_\theta^{M''}(\hat{A}^{\pm})^{2m}\,(\hat{B}^{\pm})^n
%\end{equation}
%where
%\begin{equation}
% M'' = M' \pm 2m= k\epsilon \mp 2kn \pm 2m
%\end{equation}
%Therefore, $M''= M$ when $k= m/n$, and in this case $\hat X^\pm$ are symmetries
%satisfying (\ref{symm}).
%%
Hence, once  $\sqrt{\hat{H}_{\phi}}$ is replaced by its action $\epsilon$ on
the eigenfunctions (\ref{xipm2}), the product $(\hat{A}^{\pm})^{2m}\,(\hat{B}^{\pm})^n$ will be a symmetry of
the Hamiltonian (\ref{qht4}) provided
\begin{equation}\label{kkk}
k\, \epsilon = k( \epsilon\mp 2n) \pm 2m\, .
\end{equation}
This will happen when the coefficient $k$ takes the rational value $k=m/n$.

For the case of the one--parameter PT potential, the symmetries are
obtained using (\ref{com2}) and (\ref{actionShift})
\begin{equation}\label{symmpm1}
\hat{X}^{\pm}=(\hat{A}^{\pm})^m\, (\hat{B}^{\pm})^n,\qquad k=m/n
\end{equation}
where $m, n$ are positive integer numbers. In these relations we are
making use of the simplified free--index notation. The symmetry
operators $\hat{X}^{\pm}$ are defined on the set of eigenfunctions
$\Psi(\theta,\phi)=\Theta_E^M(\theta)\,\Phi_\epsilon(\phi)$ of $\hat
H$. A comment on this question is given in the concluding section.

%%%%%%%%%%%%%%%%%%%%%%%%%%%%%%%%%%%%%%%%%%%%%%%%%%%%%%%%%55
\subsection{Degeneracy of the energy levels}

%Let us write the Hamiltonian (\ref{qht2}) in the form
%\begin{equation}\label{qht4}
%\hat{H} \equiv \hat{H} \left[k\, \sqrt{\hat{H}_{\phi}}\right]= - \partial^2_{\theta} -
%\cot{\theta}\partial_{\theta}
%+\frac1 {\sin^2 {\theta}}\left(k\, \sqrt{\hat{H}_{\phi}}\right)^2
%\end{equation}
%where we want to make explicit the dependence of $\hat{H}$ on the one dimensional
%Hamiltonian
%$\hat{H}_{\phi}$.

Along this subsection an eigenfunction separated in the variables $\theta,\phi$
corresponding to the eigenvalue $E$ will be denoted by
\begin{equation}
\Psi_E= \Theta_E^{k\epsilon}(\theta)\,\Phi_\epsilon(\phi),\qquad M= k\epsilon
\end{equation}
such that
\begin{equation}\label{hsqrt}
\hat{H}%\left[k\, \sqrt{\hat{H}_{\phi}}\right]
\Psi_E = E\, \Psi_E \, .
\end{equation}
This is satisfied, as shown in Section 2, if
\begin{equation}\label{ee}
\hat{H}_{\phi}\, \Phi_\epsilon = \epsilon^2 \,\Phi_\epsilon\, ,
\qquad
\hat{H}_\theta^M
\Theta_E^{k\epsilon}= E\, \Theta_E^{k\epsilon},\qquad  M= k\epsilon\, .
\end{equation}
%where $\hat{H}_\theta\left[k\, \epsilon\right]$ is the Hamiltonian (\ref{hsqrt})
%where $\hat{H}_{\phi}$ has been replaced by $\epsilon^2$.

If we act, for instance, with the symmetry $\hat{X}^{+}=(\hat{A}^{+})^{2m}\,(\hat{B}^{+})^n$
on this eigenfunction, according to (\ref{xipm}) and (\ref{bpmpsi}) we will get
(up to normalization constants)
\begin{equation}
(\hat{A}^{+})^{2m}\,(\hat{B}^{+})^n\, \Theta_E^{k\epsilon}\,\Phi_\epsilon=
\Theta_E^{k\epsilon+2m}\,\Phi_{\epsilon+2n} \, .
\end{equation}
Taking into account that $k=m/n$, we get
\[
k\epsilon+2m = k\epsilon',\qquad \epsilon'=\epsilon+2n \, .
\]
Therefore, the new function
$
\Theta_E^{k\epsilon+2m}\,\Phi_{\epsilon+2n}=
\Theta_E^{k\epsilon'}\,\Phi_{\epsilon'}
$,
is another eigenfunction of $\hat H$, with same eigenvalue as the initial one.
By applying $r$ times the symmetry $\hat{X}^{+}$,
we will get other eigenfunctions with the same eigenvalue:
\begin{equation}
(\hat{X}^{+})^r\,\Theta_E^{k\epsilon}\,\Phi_\epsilon=
\Theta_E^{k\epsilon+2mr}\,\Phi_{\epsilon+2nr}\, .
\end{equation}
In a similar way more eigenfunctions in the same eigenspace can be obtained by
applying $\hat{X}^{-}$,
\begin{equation}
(\hat{X}^{-})^s\,\Theta_E^{k\epsilon}\,\Phi_\epsilon=
\Theta_E^{k\epsilon-2ms}\,\Phi_{\epsilon-2ns}\, .
\end{equation}
There are some limits in the powers of the symmetries.
\begin{itemize}
\item[i)]
$\epsilon-2ns\geq \epsilon_0$. The Hamiltonian $\hat{H}_{\phi}$ has a lowest eigenvalue
$\epsilon_0^2$, so we can not decrease the eigenvalues of $\hat{H}_{\phi}$ below this
ground level.
\item[ii)]
$(k \epsilon +2mr)^2<E$. This is because
the parameter $M^2=(k \epsilon +2mr)^2$ of the PT potential in $\hat{H}_{\theta}$
can not be greater than the energy $E$.
\end{itemize}
These inequalities fix the maximum values of the powers $r$ and $s$
of the symmetries that can be applied in order to get new independent eigenfunctions in the
same eigenspace. This means that the degeneracy of each eigenvalue must be finite.
Needless to say, the same considerations apply to the one--parameter
PT case.

%%%%%%%%%%%%%%%%%%%%%%%%%%%%%%%%%%%%%%%%%%%%%%%%%%%%%%%%%55
\subsection{Energy levels}

Since the values of the energy levels can be explicitly obtained by the
factorization properties of the two component PT systems,
we can check the properties of the previous subsection.
%Let us take the eigenfunction
%$\Psi_E= \Theta^E_{k\epsilon}\,\Phi_\epsilon$ characterized by the eigenvalues
%(\ref{ee}) $E_\theta=E$ and $E_\phi= \epsilon^2$ corresponding to the
%one--dimensional Hamiltonians $\hat H_\theta^M$ and  $\hat H_\phi$ (choosing, for instance,
%the two--parameter PT case).

Each of the one--dimensional PT Hamiltonians $\hat H_\theta^M$ and $\hat H_\phi$
has the spectrum characterized by  positive integer numbers $\mu$ and $\nu$,
respectively \cite{calzada12,sengul09}, that can be written in the following notation
%In this subsection it is convenient to use the following notation
%for eigenvalues and eigenfunctions,
\begin{equation}\label{eigen}
\begin{array}{ll}
\hat H_\theta^{M}\Theta_\mu^{M}
=E_\mu^{M} \Theta_\mu^{M}\, ,
\quad &E_\mu^{M}= ({M} +\mu)({M}+\mu +1), \quad \mu = 0,1,\dots
\\[2.ex]
\hat H_\phi\Phi_\nu= (\epsilon_\nu)^2 \Phi_\nu\, ,
\quad &\epsilon_\nu = \alpha + \beta +2\nu +1,\quad
\nu= 0,1,\dots
\end{array}
\end{equation}
In this notation the separated eigenfunctions of the total Hamiltonian (\ref{qh2})
are
\begin{equation}
\hat H \,(\Theta^{k\epsilon_\nu}_\mu\Phi_\nu) =
E\,(\Theta^{k\epsilon_\nu}_\mu\Phi_\nu),\qquad M= k\epsilon_\nu\,,
\end{equation}
with eigenvalues
\begin{equation}\label{emunu}
E=E_\mu^{k\epsilon_\nu}= ({k\epsilon_\nu} +\mu)({k\epsilon_\nu}+\mu +1),\qquad
\mu,\nu= 0,1\dots
\end{equation}
Different values of $(\mu,\nu)\neq(\mu',\nu')$,  label
different eigenfunctions. However, two of the eigenvalues (\ref{emunu}) can be equal if
\[
k\epsilon_\nu +\mu = k\epsilon_{\nu'} +\mu'\, .
\]
This happens, for instance, when $k=m/n$ and
$\nu' = \nu \mp n$, $\mu' = \mu \pm 2m$,
%
%Therefore, having in mind that
%$M= k \epsilon$, the eigenvalues of the total Hamiltonian described by this type of
%eigenfunctions are labeled by two positive integers, $(\mu,\nu)$, and are
%given by the following formula
%\begin{equation}\label{emn}
%\begin{array}{l}
%E_{\mu,\nu} = (k(\alpha+\beta+2\nu+1) +\mu)(k(\alpha+\beta+2\nu+1)+\mu +1)
%\\[2.ex]
%\quad\quad  = (k(\alpha+\beta+1)+2k\nu +\mu)(k(\alpha+\beta+1)+2k\nu +\mu+1)
%\end{array}
%\end{equation}
%If $k= m/n$ the simultaneous change
%\begin{equation}
%(\mu,\nu) \to ( \mu \mp 2m, \nu \pm n)
%\end{equation}
%will leave the energy invariant:
%\[
%E_{\mu,\nu}= E_{\mu \mp 2m, \nu \pm n},\quad k=m/n\, .
%\]
which corresponds to the action of the symmetry operators $\hat
X^\pm$ on the eigenfuntions.

%%%%%%%%%%%%%%%%%%%%%%%%%%%%%
\section{The classical system}
The classical Hamiltonian function on the sphere  corresponding
to the quantum system (\ref{qh2}) %Smorodinsky--Winternitz  type systems (\ref{qh2})
has the form
\begin{equation}\label{ch2}
H = p_{\theta}^2 +\frac{p_\phi^2}{\sin^2 {\theta}}
+\frac{k^2\alpha^2}{\sin^2 {\theta}\,\cos^2{k\,\varphi}}
+\frac{k^2\beta^2}{\sin^2 {\theta}\,\sin^2{k\,\varphi}} \, ,
%=p_{\theta}^2+\frac{k^2\,H_\phi}{\sin^2 {\theta}}
\end{equation}
where the first two terms are for the kinetic energy and the remaining ones are for the potential.
We will perform a  change of the canonical variables %$(\varphi,p_\varphi)$ by
$(\phi = k\varphi,p_\phi = p_\varphi/k)$, with the same comments on the range of
$\varphi$ and $\phi$ mentioned in the quantum case. The constant $k$ is also assumed
to satisfy $k\geq 1/4$ in order that the classical system be well defined in  a region of the sphere.
In the new canonical variables, the Hamiltonian is
\begin{equation}\label{ch2}
H = p_{\theta}^2 +\frac{k^2}{\sin^2 {\theta}}\left(
p^2_{\phi}+\frac{\alpha^2}{\cos^2{\phi}}+\frac{\beta^2}{\sin^2{\phi}}\right)\, .
%=p_{\theta}^2+\frac{k^2\,H_\phi}{\sin^2 {\theta}}
\end{equation}
This classical system can also be considered as composed of two
effective one--dimensional systems $H_\phi$ and $H_\theta^M$ defined
as follows. $H_\phi$ is the two-parameter PT Hamiltonian  function
\begin{equation}\label{hc}\nonumber
H_\phi = p^2_\phi + \frac{\alpha^2}{\cos^2\phi}
+\frac{\beta^2}{\sin^2 \phi}\,.
\end{equation}
If $\beta=0$ (or $\alpha=0$), it comes into the classical one--parameter PT potential.
The second component is
\begin{equation}\label{cht1}
H_{\theta}^M=p_{\theta}^2+\frac{k^2\,H_\phi}{\sin^2
{\theta}}=p_{\theta}^2+\frac{M^2}{\sin^2 {\theta}}\,,\qquad M=k\sqrt{H_\phi}
\end{equation}
which is just another one--parameter classical PT Hamiltonian.

From the separation of variables we have an initial constant of motion $H_\phi$.
In order to get other nontrivial constants of motion we will follow the
same procedure as in the quantum  case. So, first the relevant
ladder and  shift functions will be found and, in a second stage, by combining them
the symmetries will be easily constructed.

%%%%%%%%%%%%%%%%%%%%%%%%%%%%%%%%%%%%%%%%%%%%%%%%%%%%%%55
\subsection{Ladder functions of the classical two--parameter P\"oschl--Teller Hamiltonian}
%A one-dimensional component of the total Hamiltonian (\ref{ch2}) is the
%classical two-parameter PT Hamiltonian function
%\begin{equation}\label{hc}\nonumber
%H_\phi = p^2_\phi + \frac{\alpha^2}{\cos^2\phi}
%+\frac{\beta^2}{\sin^2 \phi}\,.
%\end{equation}
%When $\beta=0$ (or $\alpha=0$), we get the classical one--parameter PT potential.
In a similar way to the ladder operators for the quantum case, we
have  two ladder functions for $H_\phi$ given by \cite{sengul08}
\begin{equation}\label{bc}
%\begin{array}{l}
B^{\pm} ={\pm}i p_\phi\,\sin 2 \phi+\sqrt{H_\phi} \cos 2
\phi+\frac{\beta^2-\alpha^2}{\sqrt{H_\phi}}
%\\[2.ex]
%\Xi_\epsilon^-=-i p_\theta\,\sin(2 k\theta)+ \sqrt{H_\theta} \cos(2
%k\theta)+\frac{\beta^2-\alpha^2}{\sqrt{H_\theta}}
%\end{array}
\end{equation}
that together with the Hamiltonian, $H_\phi$,   satisfy the
following Poisson brackets (PBs)
\begin{equation}\label{hbb}
\{H_\phi,B^{\pm}\}= \mp 4 \,i\sqrt{{H_\phi}}\,B^{\pm}\,.
\end{equation}
Remark that the canonical variables are assumed to satisfy $\{\phi,p_\phi\}=1$.
The multiplication of these
two functions gives another function depending only on the
Hamiltonian $H_\phi$,
\begin{equation}\label{bb}
B^{+}B^{-}=H_\phi+\frac{(\beta^2-\alpha^2)^2}{H_\phi}-2(\beta^2+\alpha^2)\,.
\end{equation}
As a consequence of  (\ref{hbb}), the ladder functions and ${H}$ fulfill the following PBs,
\begin{equation}\label{commpt2}
\{H ,B^{\pm}\}=\mp\frac{4\,i\,M\,k}{\sin^2 {\theta}}\,B^{\pm}\, .
\end{equation}
%where $M=k\,\sqrt{H_\phi}$.

%%%%%%%%%%%%%%%%%%%%%%%%%%%%%%%%%%%%%%%%%%%%%%%%%%%%%%55
\subsection{Ladder functions of the classical one--parameter P\"oschl--Teller Hamiltonian}
The Hamiltonian function for the one--parameter PT potential is
\begin{equation}\label{cpth}
H _{\phi}=p^2_{\phi}+\frac{\alpha^2}{\cos^2{\phi}}\,.
\end{equation}
The ladder functions for this Hamiltonian have been obtained in
\cite{sengul08}. Here, we will briefly recall how to get them. As in the quantum
 case  multiplying $H _{\phi}$ by $\cos^2{\phi}$ and
rearranging it we get
\begin{equation}\label{cpth1}
p^2_{\phi}{\cos^2{\phi}}-H
_{\phi}{\cos^2{\phi}}=B^+\,B^-+\lambda_B=-\alpha^2
\end{equation}
where
\begin{equation}\label{cbpm}
B^{\pm}=\mp i \cos{\phi}\,p _{\phi}+\sqrt{H
_{\phi}}\,\sin{\phi},\qquad\lambda_B=-H _{\phi}\,.
\end{equation}
Hence, the functions $B^{\pm}$ also factorize
the Hamiltonian in the form
\begin{equation}\label{cpth2}
H _{\phi}=B^+\,B^-+\alpha^2\,.
\end{equation}
These functions together with ${H _{\phi}}$ satisfy the
following PBs
\begin{equation}\label{commpt}
\{H _{\phi},B^{\pm}\}=\mp\,2\, i\,\sqrt{H _{\phi}}\,B^{\pm}
\end{equation}
and  therefore,  their PBs with ${H }$ will be
\begin{equation}\label{commpt1}
\{H ,B^{\pm}\}=\mp\frac{2\,i\,M\,k}{\sin^2
{\theta}}\,B^{\pm}\,.
\end{equation}

%%%%%%%%%%%%%%%%%%%%%%%%%%%%%%%%%%%%%%%%%%%%%%%%%%%%%%%%%%%%%%%
\subsection{Shift functions of the Hamiltonian $H_{\theta}^M$}
%Choosing $M^2=k^2\,H_\phi$, the Hamiltonian function (\ref{ch2}) takes the form
%\begin{equation}\label{cht1}
%H_{\theta}^M=p_{\theta}^2+\frac{k^2\,H_\phi}{\sin^2
%{\theta}}=p_{\theta}^2+\frac{M^2}{\sin^2 {\theta}}\,.
%\end{equation}
The  Hamiltonian $H_\theta^M$ is factorized  in terms of two functions,
\begin{equation}\label{cht2}
H_{\theta}^M=H_M=A^{+}\,A^{-}+\lambda_{A}
\end{equation}
where
\begin{equation}\label{capm}
A^{\pm}=\mp i\,p_{\theta}+M \cot{\theta},\qquad \lambda_{A}=M^2\,.
\end{equation}
The functions $A^{\pm}$ and $H_{\theta}^M$ obey the following PBs
\begin{equation}\label{commpt3}
\{H _{\theta}^M,A^{\pm}\}=\pm\frac{\,2 i\,M}{\sin^2{\theta}}\,A^{\pm}\,.
\end{equation}
These PBs should be compared with the corresponding
quantum commutators  of the shift operators given in (\ref{coma1}).

%%%%%%%%%%%%%%%%%%%%%%%%%%%%%%%%%%%%%%%%%%%%%%5
\subsection{Symmetries of the classical system}
A pair of  functions $X^{\pm}$  are symmetries if
their PBs with the Hamiltonian function vanish,
\begin{equation}\label{csymmetr}
\{H ,X^{\pm}\}=0\, .
\end{equation}
Taking into account the PBs (\ref{commpt2}), (\ref{commpt1}) and
(\ref{commpt3}) and for rational values $k=m/n$, it is immediate to
check that a pair of symmetries, in the two--parameter case, are
given by
\begin{equation}\label{csymmet2}
X^{\pm}=(B^{\pm})^n(A^{\pm})^{2m}
\end{equation}
and for the one--parameter case the symmetries take the form
\begin{equation}\label{csymmet1}
X^{\pm}=(B^{\pm})^n(A^{\pm})^m\,.
\end{equation}

These symmetries have  complex constant values $Q_1^{\pm}$ that (in the two--parameter
PT case) we write as follows
\begin{equation}\label{csymmet3a}
X^{\pm}=(B^{\pm})^n(A^{\pm})^{2m}=Q_1^{\pm}\equiv q_1 \,e^{\pm i \,\phi_0}
\end{equation}
%and ${\rm Re}(X^{\pm})=\rm const.$
%\begin{equation}\label{csymmet3}
%X^{\pm}=q_1\,e^{\pm i \phi_0}=Q_1^{\pm}
%\end{equation}
where the modulus of $Q_1^{\pm}$ is
\begin{equation}\label{mod}
q_1 =
\left(E_\phi+\frac{(\beta^2-\alpha^2)^2}{E_\phi}-2(\beta^2+\alpha^2)\right)^{n/2}
(E_{\theta}-k^2\,E_{\phi})^m
\end{equation}
and $e^{\pm i \,\phi_0}$ are the phases.
Besides, we should add two immediate
constants of motion: $H$ itself and $H_{\phi}$,
having values denoted by  $E_{\theta}\equiv E$ and $E_{\phi}$.
Thus, in principle we have a total of four constants of motion. However, the product
$X^+X^-$ depends only on $E_{\theta}$ and $E_{\phi}$. This follows from
the products $B^-B^+$ given in (\ref{bb}) and $A^-A^+$ in (\ref{cht2}),
so that in fact we have just only three independent
constants of motion fixed by the parameters
$E_\theta$, $E_\phi$ and $\phi_0$ (modulo $2\pi$). The constants
$E_\theta,E_\phi$ are subject, according to (\ref{mod}) to the inequalities
\begin{equation}\label{ineq}
E_\phi+\frac{(\beta^2-\alpha^2)^2}{E_\phi}-2(\beta^2+\alpha^2)>0,
\qquad
E_{\theta}>k^2\,E_{\phi}\, .
\end{equation}
In this way,  we have obtained the maximal
superintegrability of this system.

%(E_{\theta}-k^2\,E_{\phi})
%\left(E_\phi+\frac{(\beta^2-\alpha^2)^2}{E_\phi}-2(\beta^2+\alpha^2)\right)^{1/2k}
%%%%%%%%%%%%%%%%%%%%%%%%%%%%%%%%%%%%%%%%%%%%%%5
\subsection{Trajectories of the classical system}
\begin{figure}
\centering
\includegraphics[width=0.35\textwidth]{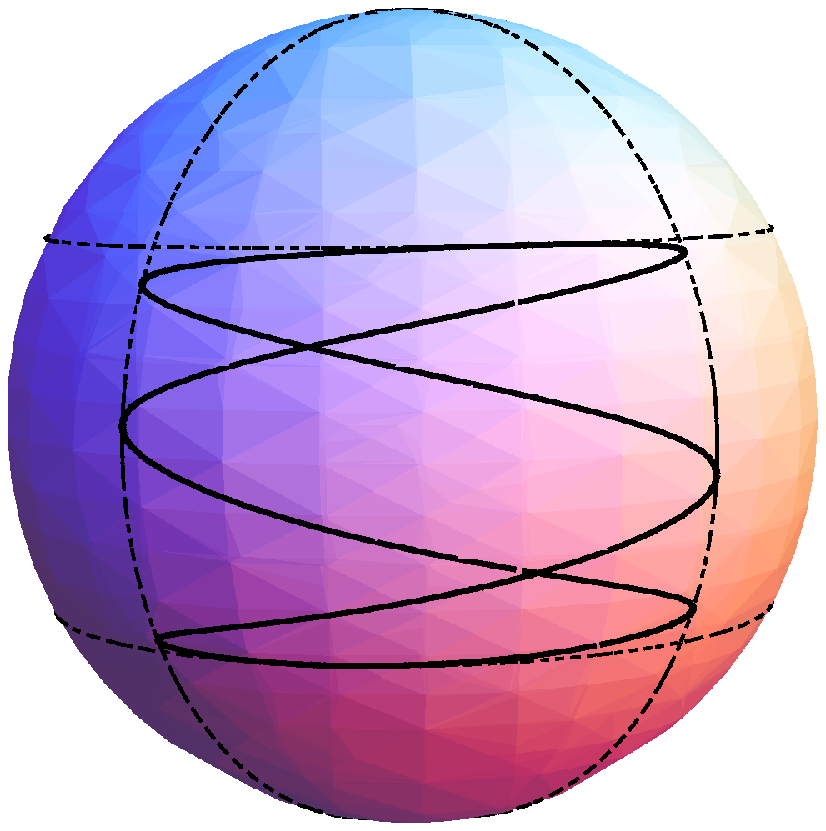}\qquad
\includegraphics[width=0.35\textwidth]{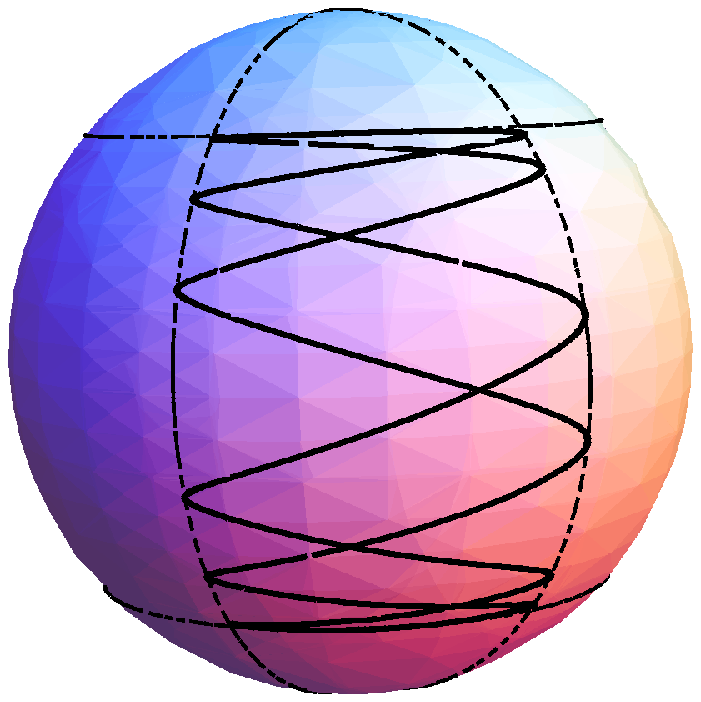}
\caption{Plot of $(\phi,\theta)$-trajectories for $k=3$. The figure
on the left is for the one--parameter PT potential $H_\phi$,
$\omega_\phi = 3\omega_\theta$. The right hand figure is for the
two--parameter PT potential $\omega_\phi = 6\omega_\theta$ (right).
These trajectories are inscribed on  `spherical rectangles' (dashing
curves). \label{k3}}
\end{figure}

\begin{figure}
\centering
\includegraphics[width=0.35\textwidth]{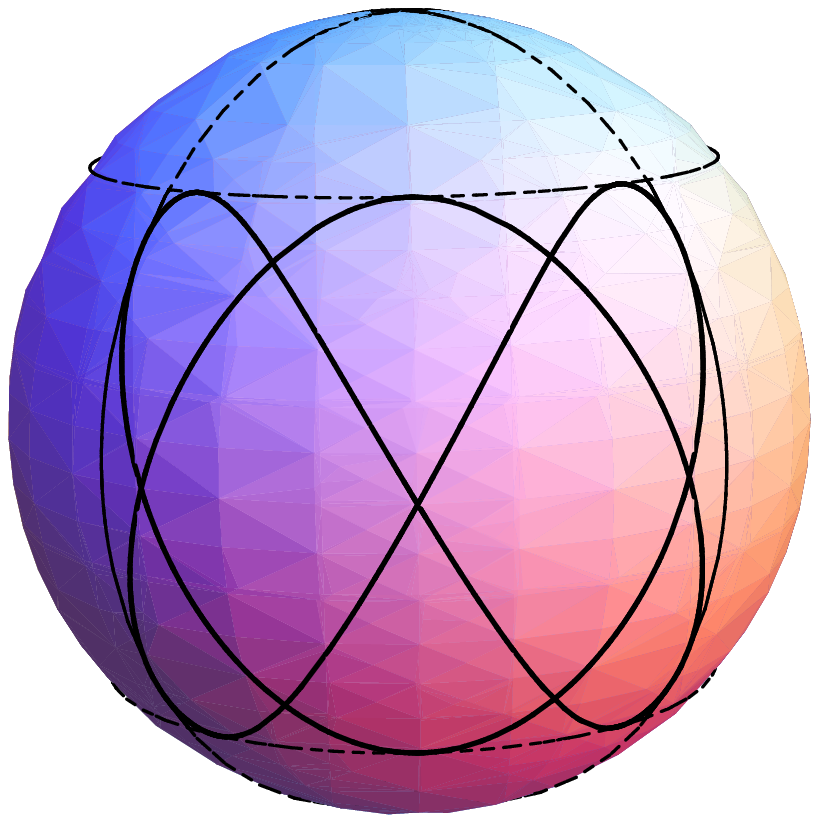}\qquad
\includegraphics[width=0.35\textwidth]{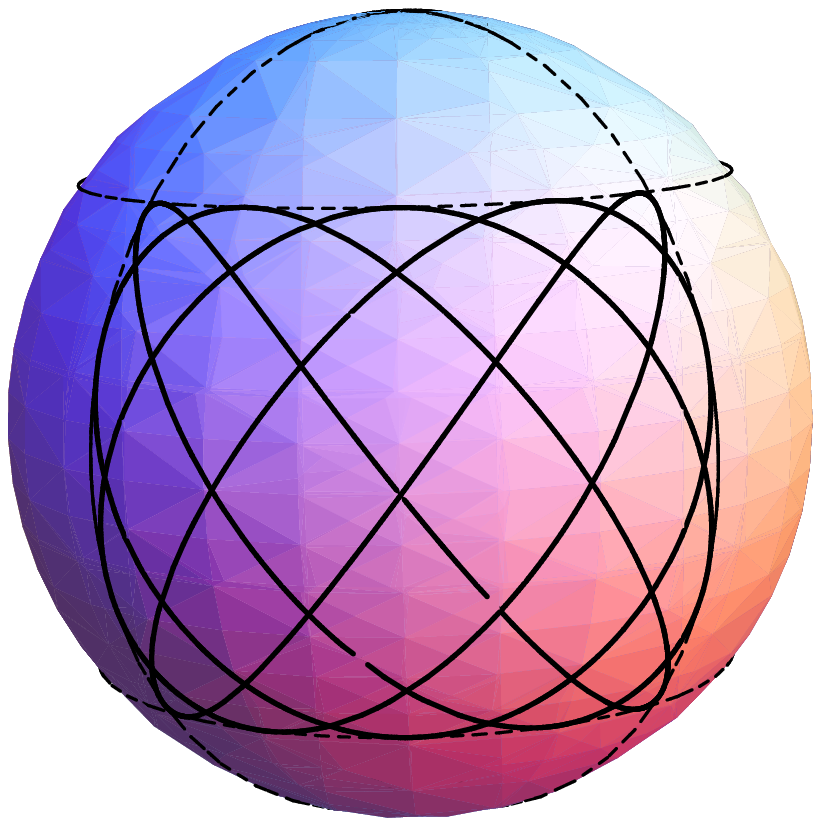}
\caption{Plot of $(\phi,\theta)$-trajectories for $k=2/3$,
$3\omega_\phi = 2\omega_\theta$ (left) and $k=4/5$,  $5\omega_\phi =
4\omega_\theta$ (right) of the one--parameter PT system inscribed in
`spherical rectangles' (dashing curves). \label{k45}}
\end{figure}

\begin{figure}
\centering
\includegraphics[width=0.35\textwidth]{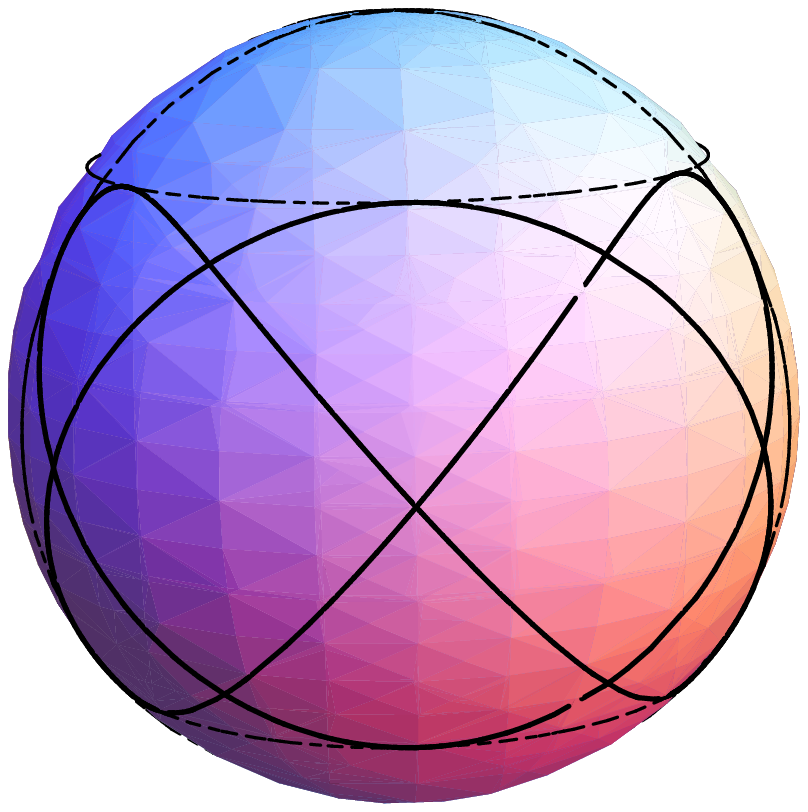}\qquad
\includegraphics[width=0.35\textwidth]{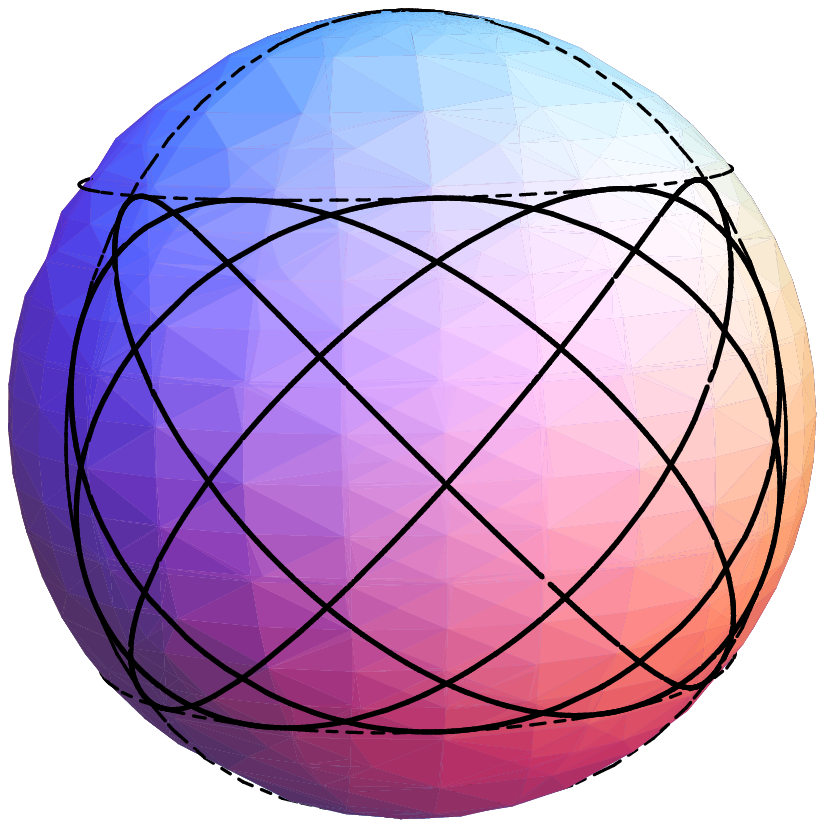}
\caption{Plot of same trajectories for $k=2/3$  (left)
and $k=4/5$ (right) as in Fig.~\ref{k45} but in
`true' spherical coordinates $(\varphi,\theta)$.
\label{k45B}}
\end{figure}

Once fixed the values of $E_\theta$ and $E_\phi$ subject to the restriction
(\ref{ineq}), the shift and ladder
functions (choosing for instance the two--parameter PT potential) can be expressed in
the form
\begin{equation}\label{fases1}
\begin{array}{l}
\displaystyle
A^\pm(\theta,p_\theta) = (E_{\theta}-k^2\,E_{\phi})^{1/2}\, e^{\pm i a(\theta,p_\theta)}
\\[2.ex]
\displaystyle
B^\pm(\phi,p_\phi) = \left(E_\phi+\frac{(\beta^2-\alpha^2)^2}{E_\phi}-2(\beta^2+\alpha^2)\right)^{1/2}e^{\pm i b(\phi,p_\phi)}
\end{array}
\end{equation}
where the real functions $a(\theta,p_\theta)$ and $b(\phi,p_\phi)$ are
identified as phase functions that can depend also
on $E_\theta$ and $E_\phi$. The Hamiltonian
$H_\theta^M$ (\ref{cht1}) describes a one--dimensional periodic system
since the potential $M^2/\sin^2\theta$ consists in an infinite well.
The values of $\theta$ are bounded by the two turning points $\theta_1,\theta_2$ determined by the equation
\begin{equation}
E_\theta= \frac{M^2}{\sin^2\theta}\, .
\end{equation}
As $(\theta,p_\theta)$ runs through a complete cycle
in the phase space corresponding to $H_\theta^M$, such that
$\theta_1\leq \theta \leq \theta_2$, the phase
$a(\theta,p_\theta)$ will increase in $2\pi$. The same will happen with
the variables $(\phi,p_\phi)$ when $\phi$ ranges between the
two turning points $\phi_1\leq \phi \leq \phi_2$ in the phase space of $H_\phi$
determined by
\begin{equation}
E_\phi=\frac{\alpha^2}{\cos^2{\phi}}+\frac{\beta^2}{\sin^2{\phi}}\, .
\end{equation}
From the symmetries $X^\pm$ and (\ref{csymmet3a})--(\ref{fases1})
the two phase functions are related by:
\begin{equation}\label{fases}
2m\,a(\theta,p_\theta)+ n\, b(\phi,p_\phi) = \phi_0\,,\qquad
\phi_1\leq \phi \leq \phi_2,\, \ \theta_1\leq\theta\leq\theta_2\, .
\end{equation}
This equation together with the constants $E_\theta, E_\phi$ fixes
the orbit (or trajectory) of the motion. In particular, for a real
motion, the variables $(\theta,\phi,p_\theta,p_\phi)$ can be
parameterized by the time and differentiating (\ref{fases}), we get
\begin{equation}
2m\,\dot a(\theta,p_\theta)+ n\, \dot b(\phi,p_\phi)=0 \, .
\end{equation}
In other words, the velocities of the two phase functions are proportional,
and therefore the frequencies (defined as the inverse of the periods)
will be also proportional (the minus sign means that the motion of the
two phases have opposite sense):
\begin{equation}\label{omega2}
2m\, \omega_\theta + n\, \omega_\phi = 0 \, .
\end{equation}
This relation is the analogue of the Lissajous curves: the periods (or frequencies)
have a rational quotient. These closed curves will be inscribed in the
`spherical rectangle' limited by $(\theta_1,\theta_2)$ and $(\phi_1,\phi_2)$.
In the case of the one--parameter PT potential this relation would be
\begin{equation}\label{omega1}
m\, \omega_\theta + n\, \omega_\phi = 0 \, .
\end{equation}
These considerations allow us to find easily the trajectories of the
system for different values of the parameters. Some examples are
shown in Figs.~\ref{k3}-\ref{k45}, where it can be appreciated that
the trajectories of the one--parameter and two--parameter PT cases
differ in the ratio of frequencies by a factor 2. The trajectories
have been given in the angles $(\phi,\theta)$, if we want to
represent the trajectories in the initial `true' spherical
coordinates $(\varphi,\theta)$, a simple dilation $\phi = k \varphi$
must be applied. The resulting graphics share the same features as
it is shown in Fig.~\ref{k45B}.

If we know  the motion in one variable (say $\theta(t)$), then the relation
(\ref{fases}) will give the motion in the other variable $\phi(t)$,
and in this way the complete motion will be determined.
This question will be discussed in the following section.

%%%%%%%%%%%%%%%%%%%%%%%%%%%%%%%%%%%%%%%%%%%%%%5
\section{Constants of motion depending explicitly on time }

It is possible to find also the motion of this system in an algebraic way
by means of the ladder functions for $H_{\theta}^M$.
These ladder functions will lead us to two
constants of motion including the time explicitly, which will allow us to
obtain the motion algebraically.
%%%%%%%%%%%%%%%%%%%%%%%%%%%%%%%%%%%%%%%%%%%%%%5
%%\subsection{Ladder functions of the Hamiltonian $H_{\theta}^M$ }

%For the Hamiltonian function $H_{\theta}^M$, its
The ladder functions of $H_{\theta}^M$ can be
found in the same way as it was shown in Section 3.2,
%First, we multiply
%$H_{\theta}^M$ by $\sin^2{\theta}$, arrange the terms, and then
%factorize the resulting function as
%\begin{equation}\label{ctd}
%-M^2=\sin^2{\theta}\,p_{\theta}^2-\sin^2{\theta}\,H_{\theta}^M=D^+D^-+\lambda_D
%\end{equation}
%where
\begin{equation}\label{dpm}
D^{\mp}=\mp
i\,\sin{\theta}\,p_{\theta}+\cos{\theta}\,\sqrt{H_{\theta}^M}\, .%,\qquad\lambda_D=-H_{\theta}^M\,.
\end{equation}
They satisfy the following PB relations with $H_{\theta}^M$,
\begin{equation}\label{commpt4}
\{H _{\theta}^M,D^{\pm}\}=\mp 2\,i\,\sqrt{H_{\theta}^M}\,D^{\pm}\,.
\end{equation}
%%%%%%%%%%%%%%%%%%%%%%%%%%%%%%%%%%%%%%%%%%%%%%5
%\subsection{Constants of motion depending on time explicitly}
Using the above ladder functions,  a set of two constants of
motion depending explicitly on time can be defined,
\begin{equation}\label{q2pm}
Q^{\pm}_2=D^{\mp}\, e^{\pm 2\,i\,\sqrt{H_{\theta}^M}\,t}
\end{equation}
such that
\begin{equation}\label{q}
\frac{dQ^{\pm}_2}{dt}=\frac{\partial Q^{\pm}_2}{\partial
t}+\{Q^{\pm}_2,H_{\theta}^M\}=0\,.
\end{equation}
These constants of motion have complex values denoted by
\begin{equation}\label{q2v}
Q^{\pm}_2=q_2\,e^{\mp i\,\theta_0}
\end{equation}
where $q_2=\sqrt{E_{\theta}-k^2\,E_{\phi}}$. Substituting
(\ref{dpm}) in (\ref{q2pm}) and using in (\ref{q2v}), $\theta$ and
$p_{\theta}$ are found  as functions  of time:
\begin{equation}\label{thetat}
\theta(t)=\arccos[\frac{q_2}{\sqrt{E_{\theta}}}\cos{(2\,\sqrt{E_{\theta}}\,t+\theta_0)}]\,,
\end{equation}
\begin{equation}\label{pthetat}
p_\theta(t)=\frac{q_2}{\sqrt{1-\cos^2{\theta}}}
\sin{(2\,\sqrt{E_{\theta}}\,t+\theta_0)}\,.
\end{equation}
The above formulas imply that the angular frequency and period of the variables
$\theta(t),p(t)$ are
\begin{equation}\label{ftheta}
\omega_\theta= 2\,\sqrt{E_{\theta}}\,, \qquad T_\theta = \pi/\sqrt{E_{\theta}} \,.
\end{equation}
Thus, the frequency comes from the bracket (\ref{commpt4}) of the ladder
functions, so it is just determined by an algebraic property of the system.
The physical meaning of (\ref{ftheta}) is that the frequency is proportional
to the square root of the total energy: the higher is the energy the
bigger will be the frequency of the periodic motion.
The frequency in the other variable $\phi(t)$ is supplied by the symmetry relation (\ref{omega1}).

%\begin{figure}
%\centering
%\includegraphics[width=0.4\textwidth]{fig_112.eps}\qquad
%\includegraphics[width=0.4\textwidth]{fig_113.eps}
%\caption{Plot of trajectories for $k=1/2$, $2\omega_\phi = \omega_\theta$ (left) and $k=1/3$,
%$3\omega_\phi = \omega_\theta$ (right) of the one-parameter PT system inscribed in rectangles. \label{k13}}
%\end{figure}

%%%%%%%%%%%%%%%%%%%%%%%%%%%%%%%%%%%%%%%%%%55
\section{Conclusions}
%%%%%%%%%%%%%%%%%%%%%%%%%%%%%%%%%%%%%%%%%%%5

The kind of quantum systems we have considered in this paper are
characterized by a separation of variables, such that
for each variable they give rise to factorizable one--dimensional systems.
The factorization properties of  these one--dimensional component systems
allow the construction of nontrivial symmetries. We have shown in our example
how this process is carried out and the way that the symmetries can be applied to
find the degeneracy of the energy levels. Although these symmetries are non
polynomial, they can directly lead us to the polynomial ones \cite{polynomial14}.

We have called Lissajous systems to the corresponding classical systems due
to similarities of their trajectories with the Lissajous curves. These
classical systems keep the same separation properties giving rise to one--dimensional
classical systems.
Although it is not well known, a whole class of classical one--dimensional systems
have analogue factorization properties as their quantum counterparts \cite{sengul08}.
This type of classical factorizations is important, for instance, in the search of
action--angle variables, or in the study of the correspondence of classical
and quantum properties through coherent states \cite{kepler12}.
By using these classical factorization properties we have shown how to get the symmetries
of the classical system in the same way as in the quantum case.
From the symmetries it is obtained the ratio of frequencies of the periodic motion in
each variable as well as the rectangles where the trajectories are inscribed.
The explicit computation of the time--dependence and the frequency of the motion
in one of the variables  is done through the ladder functions of the corresponding
one--dimensional system. In this way,  the picture of the system is complete
from an algebraic point of view.

This method to search the symmetries of classical and quantum
systems was advocated in \cite{celeghini13} as a different way to
that followed in previous references \cite{ttw09}-\cite{post}. In
general, in such references  the solutions of the Hamilton--Jacobi
equation are used in order to find the classical constants of
motion. In the quantum case, it is the solutions of the
Schr\"odinger equation which are used to get recurrence relations
and from here, the symmetry operators. We have not used any solution
at all, but only the algebraic properties of the quantum or
classical Hamiltonians.

%Therefore, the key property of a Lissajous system is that it has an
%special separable coordinate system, and this separation involves
%factorizable one--dimensional systems.
%If the superposition of the elementary one--dimensional systems is appropriate,
%then their factorization properties are responsible for the additional
%symmetries of the whole system and supplies the maximal superintegrability.

%In general, there is no more essentially different
%coordinate systems where the Lissajous systems
%can separate. This is so because the additional symmetries are
%of higher order (up to some exceptions) so that they do not give
%rise to new separable coordinate systems.
We have restricted ourselves in this paper to a very simple case
on the sphere. Our aim is just to introduce our method in the most clear way and to show
the main applications and advantages.
A thorough systematic classification of the Lissajous systems is in progress.

%%%%%%%%%%%%%%%%%%%%%%%%%%%%%%%%%%%%%%%%%%%%%%%%%%%%%%%
\subsection*{ {Acknowledgments}}
We acknowledge financial support from GIR of Mathematical Physics of
the University of Valladolid.
\c{S}.~Kuru acknowledges the warm hospitality at Department of
Theoretical Physics, University of Valladolid, Spain.

%%%%%%%%%%%%%%%%%%%%%%%%%%%%%%
%\section*{References}

\end{document}